\documentstyle[epsfig]{aipproc}

\begin{document}

\title{Extragalactic background light \\ absorption signal \\ 
in the $0.26-10$ TeV spectra of blazars.}

\author{Vladimir\ V.\ Vassiliev\thanks{Corresponding author:
vvassiliev@cfa.harvard.edu}}
\address{Whipple Observatory, Harvard-Smithsonian CfA, 
P.O. Box 97, Amado, AZ 85645, USA}
\maketitle

\vspace{-0.5cm}
\begin{abstract}
Recent observations of the TeV $\gamma$-ray spectra of the two closest
active galactic nuclei (AGNs), Markarian 501 (Mrk 501) and Markarian 421
(Mrk 421), by the Whipple and HEGRA collaborations have stimulated 
efforts to estimate or limit the spectral energy density (SED) of 
extragalactic background light (EBL) which causes attenuation of TeV 
photons via pair-production when they travel cosmological distances. 
In spite of the lack of any distinct cutoff-like feature in the spectra 
of Mrk 501 and Mrk 421 (in the interval $0.26-10$ TeV) which could 
clearly indicate the presence of such a photon absorption mechanism, 
we demonstrate that strong EBL attenuation signal (survival probability 
of $10$ TeV photon $\sim 10^{-2}$) may still be present in the spectra of 
these AGNs. This attenuation could escape detection due to a special 
form of SED of EBL and unknown intrinsic spectra of these blazars.
Here we show how the proposed and existing experiments, VERITAS, HESS, 
MAGIC, STACEE and CELESTE may be able to detect or severely limit
the EBL SED by extension of spectral measurements into the critical 
$100-300$ GeV regime.
\end{abstract}

\vspace{-0.5cm}
\subsection*{Introduction}

\begin{figure}
\begin{minipage}[t]{2.8in}
\epsfig{file=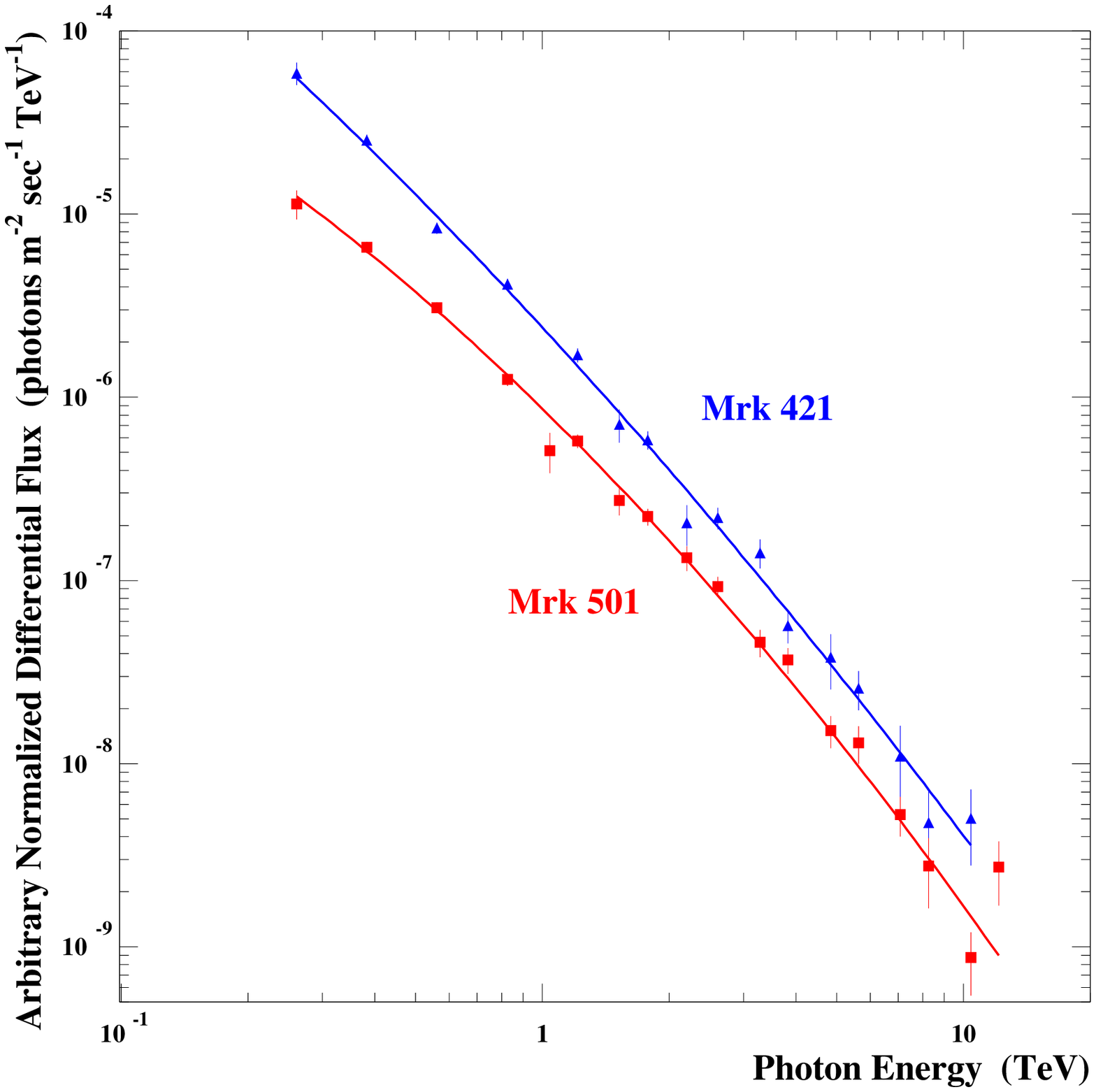,width=2.8in}
\caption{Spectra of Mrk 421 and Mrk 501 below $10$ TeV during 
high state of activity as measured by the Whipple $\gamma$-ray 
telescope [5].\label{Mrk_421_501}}
\end{minipage}
\hfill
\begin{minipage}[t]{2.8in}
\epsfig{file=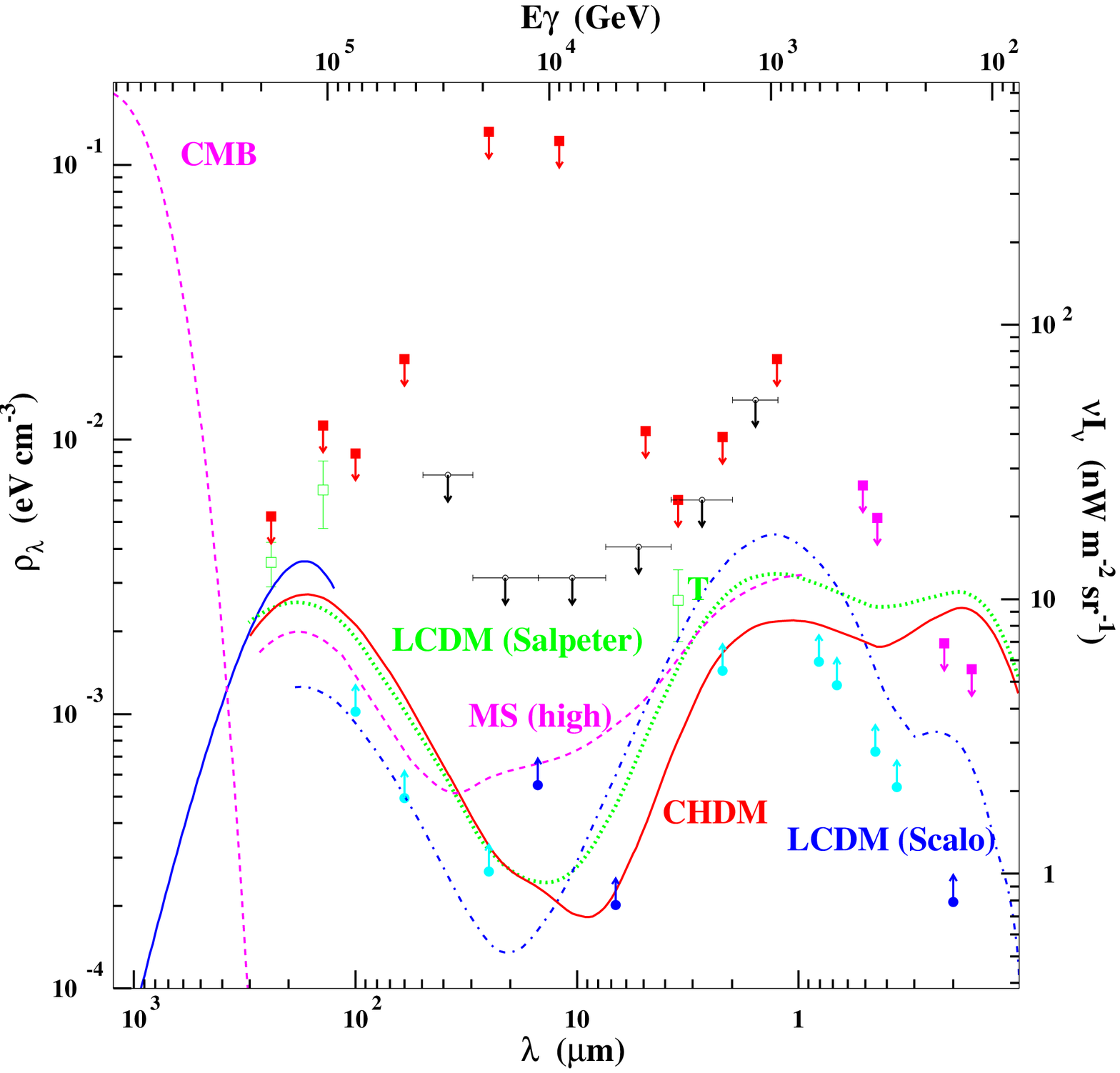,width=2.8in}
\caption{Compilation of various EBL detections, upper and lower limits, and
EBL theoretical and semi-empirical models (from [13]).
\label{ebl}}
\end{minipage}
\end{figure}

It has long been thought~\cite{GS67b} that the detection of attenuation 
effect in the TeV spectra of extragalactic sources caused by pair 
production $\gamma+\gamma \rightarrow e^+ + e^-$ with the EBL would be of 
great value for the understanding of cosmology and many aspects of the 
astrophysics of the Universe. Finding the cutoff feature in the 
high energy end of the Markarian 501 (Mrk 501) spectrum ($>10$ TeV), 
reported during this workshop \cite{Ah99}, may well be a long awaited 
signature of such extinction of the highest energy photons. 
In the presentation~\cite{Ko99} during this workshop, the claim has 
been made that the found cutoff is well explained by a semi-empirically 
derived EBL prediction~\cite{MS98} and by the simple power-law spectrum 
intrinsic to the source with an exponent close to two spanning from 
$0.2$ to $20$ TeV. The latter implies that the EBL attenuation mechanism 
below $10$ TeV should rather be weak which seems to be confirmed 
intuitively by the absence of a distinguishable feature in this part of 
the spectra for both Mrk 501 ($z=0.03$) and Mrk 421 ($z=0.03$) blazars 
(Fig.~\ref{Mrk_421_501},~\cite{Krennrich98}). We might now ask, 
whether the proposed SED of EBL is uniquely consistent with 
experiment, in order to begin its interpretation in terms of astrophysical 
constraints, or even if the lack of the feature in the low energy part 
of the TeV AGN spectra does prove an absence of the EBL absorption. Here 
I argue that we are not yet ready to make such statements neither on 
theoretical nor experimental basis, and due to ironic coincidence 
there is still a substantial degree of freedom in the definition of the 
SED of EBL. In this talk I consider a peculiar degeneracy which allows 
a certain type of SED of EBL to avoid ``apparent'' detection in currently 
available experimental data below $10$ TeV due to the unknown properties
of intrinsic spectra of the sources. The conclusions which I draw at the 
end of this talk will show how we can narrow down the existing possibilities 
even if we use spectral data of only these two AGN which may become available 
in the near future with the introduction of new $\gamma$-ray 
instruments, such as VERITAS or STACEE.

\vspace{-0.2cm}
\subsection*{The EBL}

Any contemporary theoretical consideration of the SED of EBL predicts 
two well pronounced peaks (Fig.~\ref{ebl}). One at $\sim 1$ $\mu$m due to 
the starlight emitted and redshifted through the history of the Universe, 
the other at $\sim 100$ $\mu$m generated by re-processing of the 
starlight by dust, its extinction and reemission. Theoretical modeling 
of the spectral evolution of the EBL field involve complex astrophysics 
with many unknown input parameters which specify cosmology, number density
and evolution of dark matter halos, distribution of galaxies in them, 
mechanisms of converting cold gas into stars, the star formation 
rate (SFR), stellar initial mass function (IMF), supernovae feedback,
and the mechanisms by which light is absorbed by dust and 
reemitted at longer wavelengths~\cite{SP98}. Semi-Analytical modeling 
of these processes show that the dominant factors shaping SED of EBL 
in the region $1-10$ $\mu$m are IMF, which provides a source of the 
UV light, generated mostly by the high mass stars and is therefore dependent 
on their fraction, and dust extinction which functions as a sink of 
UV light~\cite{Bu98}. The region from $1$ to $10$ $\mu$m is primarily
determined by the type of cosmology and the SFR. Allowing partial degeneracy 
between these two factors leads to an ambiguity in the interpretation of 
the actual SED of EBL~\cite{MP96}. It is also possible that a non-negligible 
contribution from a number of pregalactic and protogalactic sources may exist 
in this interval~\cite{BCH86}. This EBL fraction is usually not considered 
in the EBL evolution models nor in semi-empirical EBL estimates.  
The long wavelength region, from approximately a few $\mu$m to $100$ $\mu$m,
is currently predicted with the largest uncertainty, due to poorly defined
dust extinction and re-radiation mechanisms, which are crucial ingredients
for modeling EBL in this band. In addition, it has been suggested
recently~\cite{Fa99}, that a substantial energy in this wavelength interval 
may come from quasars. Most of their radiation should be absorbed in the
dust and gas of the accretion disk and re-emitted later in the 
far-infrared. Up to now this contribution has been considered as 
negligible, but failure to explain existing X-ray background suggests
a presence of a large population of the faint quasars generating
this diffuse radiation field~\cite{WF99}. These sources, visible only 
in X-ray and far-infrared, are expected to be probed by the Chandra mission.  
Energy ejected into the surrounding media by supernovae and re-radiated 
later may also be concealed in this wavelength interval~\cite{PB99}.

\begin{figure}
\begin{minipage}[t]{2.8in}
\epsfig{file=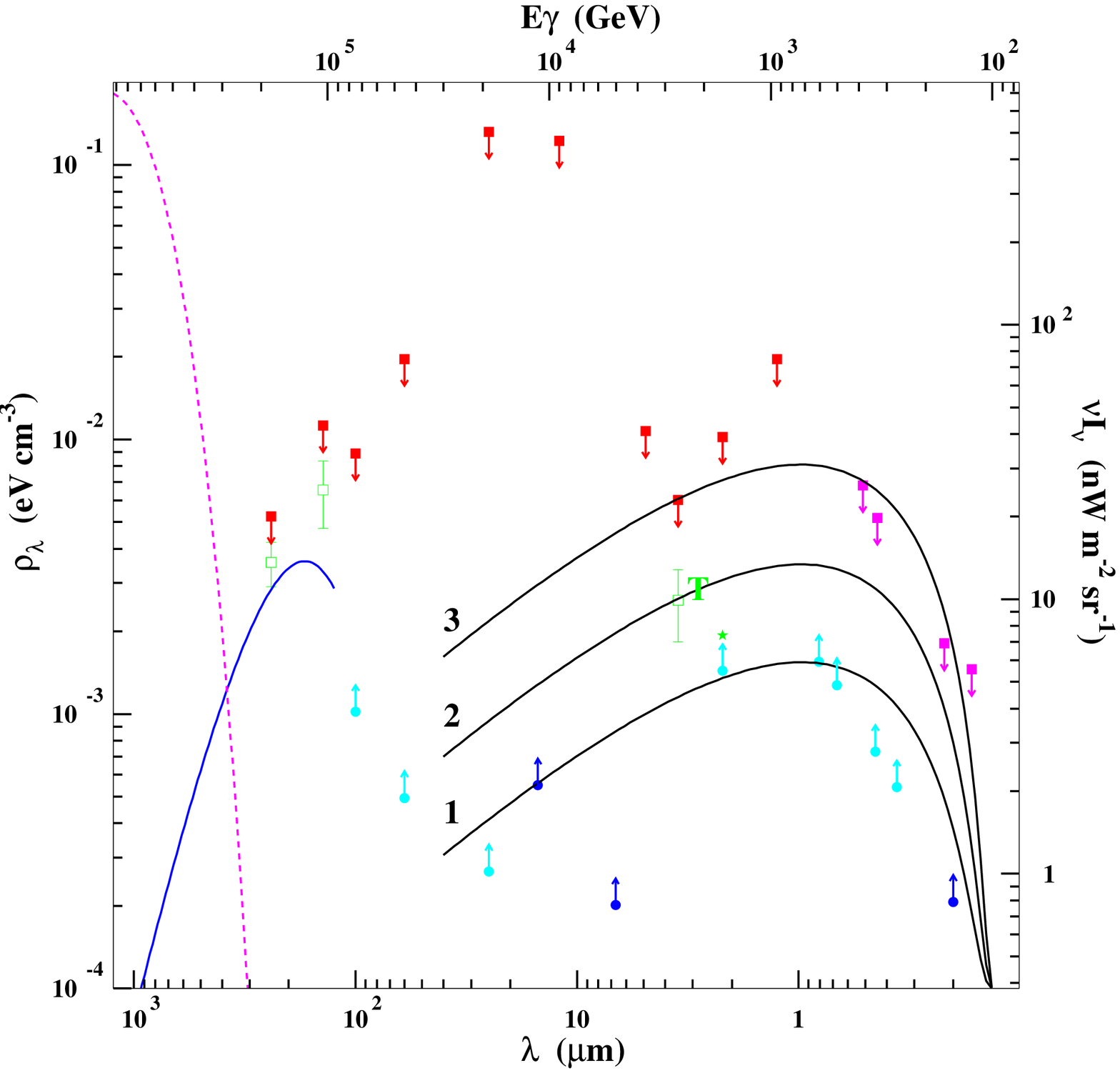,width=2.8in}
\caption{Three examples of ``invisible'' SED of EBL are 
marked 1, 2, and 3 [13].\label{solutions}}
\end{minipage}
\hfill
\begin{minipage}[t]{2.8in}
\epsfig{file=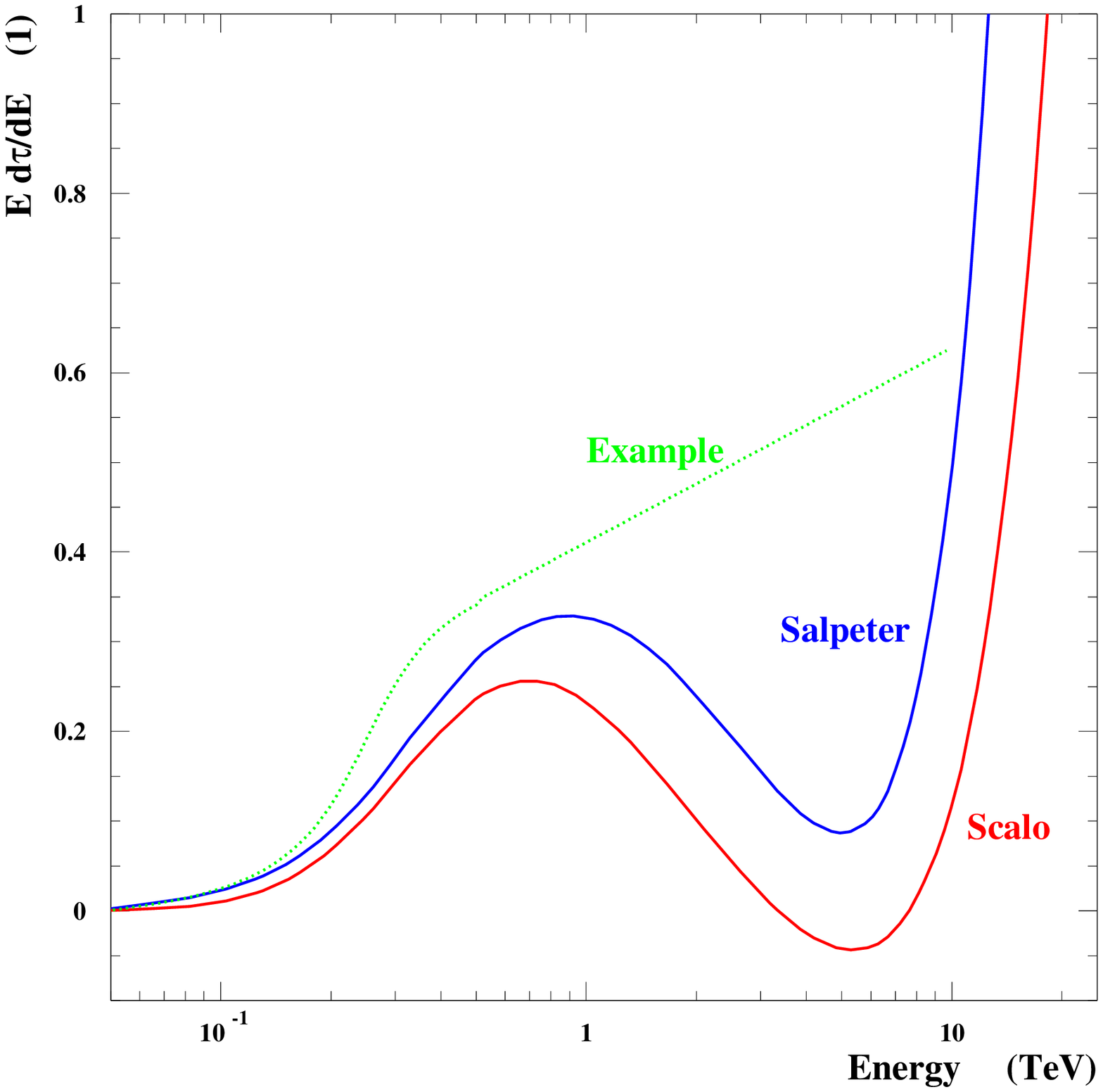,width=2.8in}
\caption{Predicted change of the Mrk 421 and Mrk 501 power-law spectral 
index due to attenuation of TeV photons on EBL [12].
\label{jump}}
\end{minipage}
\end{figure}

    At present, the degree of the uncertainty of various theoretical 
considerations of the SED of EBL is about the same as the distance between 
current upper and lower experimental limits. Fig~\ref{ebl} shows 
a compilation~\cite{VV99} of various EBL detections and limits as well as
several theoretical~\cite{PB99} and semi-empirical~\cite{MS98} estimates of 
the SED of EBL. In the current situation a preferential choice of a particular
prediction based solely on a theoretical background seems unjustified. We do 
expect, however, that the SED of EBL is likely to be a function with complex
behavior. If we take into account that attenuation of extragalactic 
TeV $\gamma$-rays is an exponential effect, one would intuitively expect
appearance of the structures in the observable spectra of AGNs, such as
cutoff, for example. Non-existence of any peculiar features in $0.25-10$ 
TeV spectra of AGNs should then indicate a very weak absorption effect. 
The problem, however, is more subtle than it first appears. There is 
a whole class of non-trivial solutions for the SED of EBL (see~\cite{VV99}), 
shown in Fig.~\ref{solutions}, which may well describe starlight peak 
expected in the $0.1-10$ $\mu$m region. The important property of such
SEDs is that they do not produce any peculiar change in the observable 
AGN spectrum. The only effect to be seen is change of the overall attenuation 
factor, change of the power-law spectral index, and slight change of the 
spectral curvature. All three of these potentially detectable EBL 
indicators are perfectly masked by the unknown intrinsic spectrum of 
the source. The existence of such SED solutions, which have been hinted 
in~\cite{DS94}, is due to slow, power-law-like, change of an attenuation 
coefficient when the SED is proportional to energy of the infrared
photon with logarithmic accuracy. Such a case seems to take place  
in the $1-10$ $\mu$m region to which observations in $0.25-10$ 
TeV interval are most sensitive. It is an ironic coincidence that 
current observational window of TeV $\gamma$-ray astronomy 
coincided with the region of a special behavior of SED which makes
EBL attenuation effect ``invisible.'' If we were to move $\gamma$-ray 
observational window to lower or higher energies we would be sensitive 
to the bands in the EBL spectrum where the SED is rapidly falling. This would 
produce an exponential effect on the observable spectra of AGNs, such as 
one detected by the HEGRA collaboration in the region $10-25$ 
GeV~\cite{Ah99}.

Since there are a number of upper and lower limits for the SED of EBL in 
the $0.1-10$ $\mu$m interval, established by various experiments,
it becomes possible to constrain the EBL attenuation effect in the spectra 
of Mrk 421 and Mrk 501 using an explicit form of ``invisible'' SEDs.
Such considerations~\cite{VV99} lead to the conclusions that the optical 
depth for $10$ TeV photons is bounded within the interval $0.85-4.43$,
the EBL contribution to the power-law spectral index at photon energy 
$1$ TeV can be in the range $0.19-0.94$, and spectral curvature does 
not exceed $0.22$ (the Hubble constant used is 65 km s$^{-1}$ Mpc$^{-1}$).
Analogous constraints derived from the EGRET measurements of the spectral 
indices of these sources~\cite{Hart99,Kt99} provide similar upper limit
for spectral index change due to the EBL attenuation effect ($<1.0$).
Source luminosity arguments~\cite{Ah99} suggest that if the optical depth
at $1$ TeV exceeds $\sim 3$ then the intrinsic $\gamma$-ray luminosity 
of Mrk 501 is an order of magnitude larger than the luminosity in all 
other wavelengths which is difficult to explain with realistic 
parameters of the jet. This places approximately the same upper bound on the
absolute value of the $\gamma$-ray absorption. Finally, a lack of the 
curvature in the spectrum of Mrk 421 provides upper limit of $0.3$
which is no stronger at the moment than the direct EBL constraints.

The large degree of the uncertainty which still exists in an experimental
detection of EBL signal via observations of $\gamma$-ray attenuation
is due to the, as yet, undetermined initial conditions in the region 
approximately from 100 to 300 GeV. Although EBL induced attenuation 
of $\gamma$-rays in this interval is rather small for z=0.03, the 
absorption effect must produce a jump in the spectral index and  
curvature here. In Fig.~\ref{jump} I show a derivative of 
the optical depth with respect to the logarithm of energy, which 
characterizes the change of the spectral index. Two examples shown correspond 
to the predictions made in~\cite{PB99} for Salpeter and Scalo IMFs.
Note that the region from 100 GeV to ~500 GeV is characterized by a 
rapid change in the spectral index. This result is, of course, hardly 
surprising since it is produced by a rapid fall of the SED of EBL in 
UV band. The subsequent dip above $1$ TeV is due to a very low 
prediction of the models for EBL field in the region around $10$ 
$\mu$m. Such a low estimate is not currently supported by the lower bound 
on EBL from ISO measurements at $15$ $\mu$m~\cite{Ol97}. Absence
of the EBL signature in the spectra of Mrk 421 and Mrk 501 in the 
$0.3-10$ TeV interval suggests that the behavior of the derivative of 
optical depth should likely be linear function of the logarithm of 
photon energy (curve marked ``Example'' in Fig.~\ref{jump}).
The measurements of the AGN spectra in the region $100-300$ GeV
are of crucial importance then to experimentally determine two parameters 
of this curve and therefore unfold a unique SED "invisible" 
solution~\cite{VV99}. The expected change of spectral index, $0.19-0.94$,
of Mrk 501 and Mrk 421 between $100$ GeV and $1$ TeV due to the EBL 
absorption is a  measurable effect. If the opacity of the Universe 
to TeV $\gamma$-rays is large, most of this change should occur below 
$300$ GeV since such an effect has not been detected in the AGN spectra
above this energy~\cite{Krennrich98}. This would produce a ``knee-like'' 
feature in the spectra of these blazars in the $100-300$ GeV interval. 
If the attenuation effect is small though, it is possible that the derivative 
of optical depth remains the same linear function in this energy band 
generating only a logarithmically small EBL footprint in the AGN spectra.
In the latter case, a small curvature, $1/2 \ d^2\tau/d\ ln(E)^2 \sim 0.1$,  
in $0.1-1$ TeV energy range would be the only indicator of EBL presence.

\vspace{-0.3cm}
\subsection*{Conclusions}
{\bf 1.}The featureless spectra of the two closest AGNs in the $0.25-10$ TeV
energy band does not guarantee a low attenuation of TeV $\gamma$-rays 
via pair-production with EBL. The large class of ``invisible'' SED solutions
exists~\cite{VV99} which change only overall attenuation factor, spectral 
index, and spectral curvature of the observable AGN spectrum. Behavior 
of these SEDs is consistent with the theoretically expected starlight EBL peak
at $1$ $\mu$m, but due to the unknown intrinsic properties of the sources 
such an attenuation effect cannot be unambiguously isolated based
only on the data from this energy interval. The EBL spectral density
suggested in~\cite{Ko99} for explanation of spectral properties of 
Mrk 501 is, therefore, only one of many possibilities.

{\bf 2.}Even by observing only two known extragalactic TeV sources,
Mrk 501 and Mrk 421, we can hope to constrain or possibly detect 
SED of EBL if spectral measurements are extended into the $100-300$
GeV region where change of the spectral index would indicate a
turn on of the absorption effect. Detection of this feature by
future $\gamma$-ray observatories, for example VERITAS or STACEE, 
will provide a missing piece of information for proper unfolding of 
the SED of EBL in the wavelength interval above $0.1$ $\mu$m. 
Of course, a certain ambiguity of spectra interpretation due to 
unknown properties of the sources will remain, the presence 
of a similar feature, which is static in time, in both spectra may 
allow to disentangle or severely limit intergalactic $\gamma - \gamma$ 
extinction effect.

{\bf 3.}It turns out that the most important constraints 
for limiting or unfolding the SED of EBL may be provided
by accurate measurements of the curvature (better than $\sim 0.1$
per decade in energy) in the spectrum of Mrk 421 through the 
interval $250-10$ TeV. At the moment the spectrum of this source 
during high flaring state has been found consistent with a pure 
power-law~\cite{Krennrich98}. If the source cooperates, an 
improved statistics may give EBL upper limits in the region 
around a few micrometers lower than the currently available {\it DIRBE}
results.

\subsection*{Acknowledgments}
I thank Whipple collaboration, J. Bullock and J. Primack for providing 
data, T. Weekes and S. Fegan for valuable discussions and invaluable help.
This work was supported by grants from the U.S. Department of Energy.

\end{document}